\documentclass{ws-acs}
\usepackage{url}
\usepackage{graphics}
\usepackage{graphicx,t1enc}
\usepackage{amssymb,amsmath}

\begin{document}


\markboth{M. N Kuperman}{A model for the emergence of social
organization in primates}

\title{A model for the emergence of social organization in primates}

\author{M. N. Kuperman}

\address{Consejo Nacional de Investigaciones Cient\'{\i}ficas y
T\'ecnicas \\ Centro At\'omico Bariloche and Instituto Balseiro,
8400 Bariloche, R\'{\i}o Negro, Argentina \\
kuperman@cab.cnea.gov.ar}

\maketitle

\begin{abstract}
Recent studies have established an apparent relationship between
the repertoire of signals used for communication and neocortex
size of different species of primates and the topology of the
social network formed by  the interactions between  individuals.
Inspired by these results, we have developed a model that
qualitatively reproduces these observations. The model presents
the social organization as a self organized processes where the
size of the repertoire in one case and of the neocortex in another
play a highly relevant role.
\end{abstract}



\section{Introduction}

The social organization of animal societies exhibits a great
diversity of structures ranging from simple aggregates to
complicated, hierarchical structures. The most complex
organization is reflected in societies in which there are multiple
roles, specialization, and even a division of labor. How these
alternative structures arise, and why, remains poorly understood.
A long-standing hypothesis is that complexity at the
organizational level depends on complexity of signaling dynamics
at the individual level ---it has been argued, for example, that
human language played a critical role in the emergence of human
cultural complexity by fundamentally changing how information can
be transmitted from individual to individual and
intergenerationally \cite{may}. This hypothesis assumes that
semantic richness drives complexity at higher levels. Yet we also
observe variation in complexity across animal societies
\cite{Blum,Com,Thie} with little semantic richness. The signaling
systems present in these societies are relatively simple with, for
example, an average of approximately 16 signals and a maximum of
38 (considering all signals including those not used in the
context of social bonding) in the 42 primate species for which
data have been compiled \cite{Com}. The variation in these
societies occurs at the individual level in appropriate contextual
usage, signal form, signal repertoire size, and signaling
frequency \cite{Maes,Preu,Sey,Jan,Pol}.  This variability, which
results from individual differences in cognition, learning, social
experience, is amplified by the presence of individuals at many
different developmental stages.

On the other hand,  some works suggest that the evolution of
sociality drives the evolution of communicative complexity
\cite{Mar,Dun1,Pin}. Although both of these hypotheses are to some
extent supported by comparative and experimental data
\cite{Blum,Com,Maes,Fre}, and it seems likely that communicative
and social complexity co-evolve, the relationship between the two
remains poorly understood and has yet to be formally modelled.

Evolutionary increases in size of vocal repertoire were associated
with an increment in group size (mammals in general
\cite{Com,Kudo2001}, see Fig 1. for non humans primate species)
and time spend grooming \cite{Com}. At the same time various
measures of brain size have been positively correlated with
repertoire's size, as well as feeding innovation, learning and
tool use in birds and primates, social complexity in birds,
primates, carnivores, and some insectivores, dietary complexity in
primates and unpredictability of environments in hominid, see
\cite{Marino2005} and references therein. Furthermore, the
interplay between relative neocortex size and social organization
has been established and discussed in many works
\cite{dum1,dum2,dum3,aie,bart,jof,saw}.

Based on the ideas discussed above,  in this work we explore the
dynamics of an evolving social network composed of interacting
individuals. Previous to a detailed explanation of the model we
will make a brief description of the dynamical process. We
consider a population of individuals whose social interaction is
well represented by a network. Two individuals can interact only
if they are socially linked though the social bound alone is not
enough for a lasting interaction. In order to preserve the social
bound the individuals must establish an effective communication,
which not only will nourish and reinforce the existing link but
also trigger new communication channels. If two linked individuals
fail to communicate the may choose to break that bound and try to
establish new links with more affine individuals.

The goal of this work is to explore how the ability to
successfully communicate affects the construction of a social
organization. Specifically we want to model how the effectiveness
of the established communication affects the overall structure of
this social organization, leading to fragmentation into smaller
subgroups or isolation of some individuals. Topologically this
translates into analyzing the size of connected components in the
resulting social network by determining who interacts with whom.

The ability of the individuals to successfully communicate will be
associated to two different but complementary characteristics  of
the population. Based on some of the results presented above,
first we explore the resulting social organization considering
populations with different signal repertoire sizes and we trace
the correlation between group and repertoire sizes. Then, we
propose associating the success of the communication to the
relative neocortex size of the individuals, assuming that the
probability of an effective communication will be proportional to
the neocortex size. Once more, our assumptions are based on
previous works that have established the links between social
complexity, neocortical enlargement, and the size of social
groupings.  There are evolutionary \cite{dum2,gord,herr,byr} and
physiological \cite{saw,brot,par} evidences that support the
correlation between brain development and communication skills.

\begin{figure}[!h]
\centering
\resizebox{9cm}{!}{\rotatebox[origin=c]{0}{\includegraphics{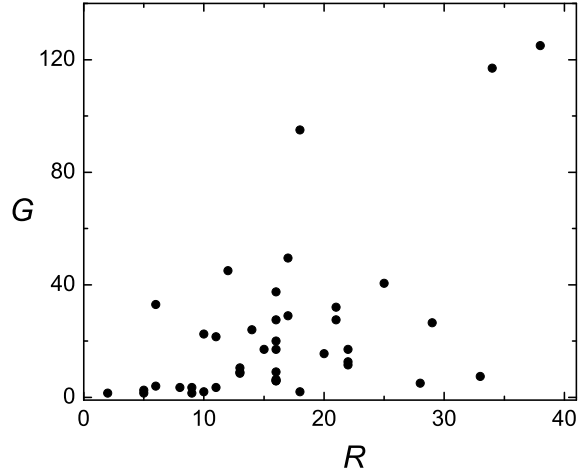}}}
\caption{Group size vs. repertory size among non humans primate
species, from \cite{Com}.} \label{staxbr}
\end{figure}

The model presented here is a generalization  of a previous set of
models exploring the causes of cultural diversity
\cite{axel1,axel,palm2,palm1,klemm,kup}. In these models culture
is defined as a set of individual attributes that are susceptible
to social influence. The individuals, placed in the nodes of a
network have attributes defined by a set of $R$ features, each
adopting one of $q$ possible traits and represented by a vector
profile of $R$ components.  The social dynamics is based on the
premise the more alike an agent is to a neighbor, the more
probably the agent will imitate one of the neighbor' s traits
\cite{axel}. The former is translated into a probability of
interaction that is  proportional to the number of common features
shared by the individuals. The interaction is imitative, leading
the two interacting individuals to the the adoption of a common
trait in one of the $R$ cultural aspects previously not shared. At
first sight this dynamics tends to homogenize the vector profile
of neighboring agents to finally converge to a monocultural state,
where all the individuals share the same vector profile
\cite{palm2,palm1,klemm,kup}. However, it has been shown that the
system can remain frozen in a multicultural states with a number
of different stable homogeneous or monocultural clusters. The
number of clusters is a measure of cultural diversity of the
steady final state.

We have adapted this model to study how variation in signal
repertoire and neocortex size may affect the social organization
in primates. In addition to the imitation dynamics analyzed in
previous works, the present model considers an evolving network
that on one side defines the interaction channels between
individuals and on the other side is simultaneously being shaped
by the distribution of repertoires in the population. The network
dynamics compete with the signaling dynamics by inhibiting the
paths to a homogeneous state, thereby producing modular structures
comprised of multiple connected components of individuals with
identical signal repertoire profiles. The coevolution phenomenon
produced by the feedback between the network topology and the
individual profiles has not been studied before and is what
defines the final social organization of the population.

We will analyze the effect of different repertoire and neocortex
sizes on the communication between individuals through an abstract
representation. Analogously to \cite{axel1} the repertoire will be
mimicked through a vector of size $R$. The vector represents the
repertoire of the individuals. Each of the components of the
vector will be associated to something that needs to be
communicated and that with some abuse of notation we will call
concept. On the other hand, the election that each individual
makes to communicate a given concept  must not necessarily be
shared by the rest of the population, thus each of the components
of the vector can adopt different values. Two individuals will
interact only if they can communicate and this implies sharing a
given number of pairs concept-signal in their respective
repertoires. In terms of the model, that means that some of the
values adopted by the same components in their respective signal
vector must coincide. The interaction in turn promotes the
imitation, generating a feedback process where a successful
interaction promotes signaling imitation, while similar repertoire
vectors enhance the interaction. As mentioned before, first we
will consider populations with different vector sizes, i.e.
different repertoire sizes.  In a second approach we will not
include the differences in repertoire sizes explicitly but we will
assume that communication is more effective among individuals with
bigger neocortex size \cite{dum2,saw,gord,byr,brot}. The basis of
this affirmation is associated to the fact that individuals with
bigger brains develop stronger communicative skills, with
increased capability to interpret social signals
\cite{Dun1,herr,par,rea}. The probability of interaction will
still  be proportional to the similarity between the vector
profiles but the effectiveness of the communication between two
individuals will be modulated by different functional forms
encoding the information about the neocortex size.

\section{The Model}

The model considers a network of interacting individuals,
initially connected through links to $2k$ neighbors. Each
individual $i$ is characterized by a repertoire profile,
represented by a vector $\vec{\rho}_i$ of $R$ components or
aspects (repertoire size). Each of the components of the
repertoire is associated to a concept or fact that needs to be
communicated. The communication operates via signaling. The
richness and variety of the used signals are associated to the
values adopted by each aspect (component) of the vector, that are
discrete values ranging between 1 and $q$ (number of traits). The
use of numbers is only a convention inasmuch they act only as
flags or labels. The repertoire profile is thus defined by the set
of $R$ values in the components of the vector $\vec{\rho}_i$.
Initially, the values adopted by each of the components of
$\vec{\rho}_i$ are randomly assigned.

As mentioned in the introduction, the interaction between the
individuals affects both the topology of the network and the
distribution of the repertoire profiles. On one side, we adopt an
imitation dynamics as a representation of the interaction that
occurs between the individuals in a population that, by favoring
the local convergence, drives the system to a uniform state with
all the individuals sharing the same repertoire profile. On the
other side, we include a dynamics that affects the network
topology and tends to freeze local inhomogeneities that may arise
by breaking links between individuals with different enough
profiles. The competition between both effects leads the system to
a self organized structure with separated groups or communities,
each one comprising individuals sharing the same repertoire
profile. A detailed explanation of both dynamics is included
below.

\subsection{Repertoire evolution}

Two randomly chosen individuals connected by a link may interact
between them with a probability $\sigma$ associated to the
similarity between their respective repertoire profiles
\cite{axel1}. Before describing the nature of the interaction, we
explain how we measure the probability of it to occur.

The similarity between two profiles is measured through the
distance  between them. In this work we present the results
obtained when considering the Hamming distance. Other metrics have
been tested producing qualitatively the same results. The Hamming
distance can be calculated as follows \cite{hamming}. Consider two
individuals $i$ and $j$, with $\vec{\rho}_i$ and $\vec{\rho}_j$
their respective repertoire profiles. Each of the $R$ components
of the repertoire profile is characterized by a value ${\rho}_i^k$
and ${\rho}_j^k$ respectively, where $k$ is the vector component
index. After counting the number of coincidences, i.e.
${\rho}_i^k={\rho}_j^k$ we obtain the Hamming distance $d_h(i,j)$
as ($R$ - number of coincidences). In turn, we need to define the
probability of interaction $\sigma$ as a function of the distance.
\begin{equation}
\sigma= 1-\frac{d_h}{R}. \label{pint}
\end{equation}

Once defined the probability of interaction, we proceed to
describe how the interaction between two individuals $i$ and $j$
affects their repertoire profiles. When the interaction is
effective one of the features $\rho^i_k$ such that $\rho^i_k \neq
\rho^j_k$ is set equal to $\rho^j_k$. That is, the individual $i$
imitates or adopts  a repertoire value already adopted by $j$ in a
randomly chosen aspect. The roles of imitator-imitated are also
randomly assigned to the pair $i$-$j$. Though it is evident that
the interaction tends to reduce the distance between
$\vec{\rho}_i$ and $\vec{\rho}_j$ it may affect  the distances
between $i$ and the rest of the neighborhood as well. First we
will consider that the communicational skills of a population are
characterized by the size of the vector profiles.

In a second version of our model we will fix the vector size and
modify Eq.(\ref{pint}) as follows

\begin{equation}
\sigma= 1-\left ( \frac{d_h}{R}\right )^{-\chi}
\end{equation}
where $\chi$ is a parameter in the interval (0,1) whose decrease
is associated to an increase in the te effectiveness of
communication. We assume that this is an indirect measure of the
effect of the neocortex size. The net effect is that a bigger
neocortex will facilitate the communication and interaction
between individuals. This fact is reflected in that the
probability of interaction has lower thresholds values for lower
values of $\chi$, i.e larger neocortex.

\subsection{Network evolution}

As mentioned above, initially the individuals are located in a
regular ordered network, with $2k$ neighbors \cite{watts}. Though
the repertoire evolution favors the local convergence to a uniform
state, the initial disordered distribution of the repertoire
profiles can lead to a situation when the distance between the
vector of a chosen subject and one of its neighbors is greater
than the distance between the vector of the subject and that of
another not neighboring one. If such is the case, the chosen
subject may prefer including the more similar or affine individual
into its neighborhood even at the cost of breaking an already
established but non interesting link (we preserve the total number
of links). This translates into the fact that links between non
similar individuals can be broken to allow the creation of links
between non connected similar individuals.

Applying the ideas discussed above, the network evolution acts in
the following way. We choose a couple of linked individuals $i$
and $j$, and a third one $k$ not linked to $i$. We measure the
distances between the repertoire vectors $d(i,k)$ and $d(i,j)$. If
$d(i,k) < d(i,j)$ then the link between $i$ and $j$ is broken
while a new link between $i$ and $k$ is established.

It is possible to understand now the competition between the
convergence and segregation, associated to repertoire and network
dynamics respectively. In our model, both dynamics  act on the
system in alternate turns. We define as $t_r$ the time length of
the repertoire dynamics turn and as $t_n$ the corresponding to
that of the network dynamics. This means that we first consider $N
\times t_r$ pair interactions, with eventual changes in the
repertoire, followed by $N \times t_n$ proposals of changes to the
topology of the network. Some of the proposed changes in the
repertoire vector and in the topology will be rejected according
to the defined dynamics.  The values $t_r$ and $t_n$ will play a
determinant role on the dynamics of he system, as will be shown
later.

\section{Results}

In the following section we describe the numerical results
obtained when considering networks of  $10^2$ and $10^3$
individuals and $k=2$, without losing generality.

\subsection{Variable vector size}

The starting network is a regular ordered one. Initially, each
individual is assigned a random repertoire vector $\vec{\rho}_i$
of dimension $2\leq R \leq 20$. The repertoire vector takes into
account the several possibilities for referring a given object or
activity that can be signaling in the community. Each component of
the vector can adopt any integer number between 1 and $q$. In the
following calculations $10\leq q \leq 50$.

Throughout the calculations we use asynchronous update. The
simulation proceeds in the following way. Once chosen the set of
parameters ${R,q}$ we need to choose the values $t_r$ and $t_n$.
The choice of these values will dictate the behavior of the system
that will range from detailed segregation into small groups or
convergence to a connected homogeneous population. When
considering the repertoire evolution each evolution unit step
consists in both cases in $N$ interactions of a randomly chosen
individual and a randomly chosen neighbor. On the other hand, when
we consider the network evolution a third randomly chosen non
neighboring individual is also picked up.  Throughout the
simulations we have taken  $1 \leq t_n \leq 100$ and $0.1 \leq t_r
\leq 10$. The characteristic time for repertoire adjustment must
be  shorter than that of network reconfiguration to avoid
stimulating a segregation process that will end in a network
composed by  isolated and non interacting small groups.

We consider that the system has reached a stationary state when
the number of changes of any nature, vector profile or network
topology, goes to zero.

\begin{figure}[!h]
\centering
\resizebox{9cm}{!}{\rotatebox[origin=c]{0}{\includegraphics{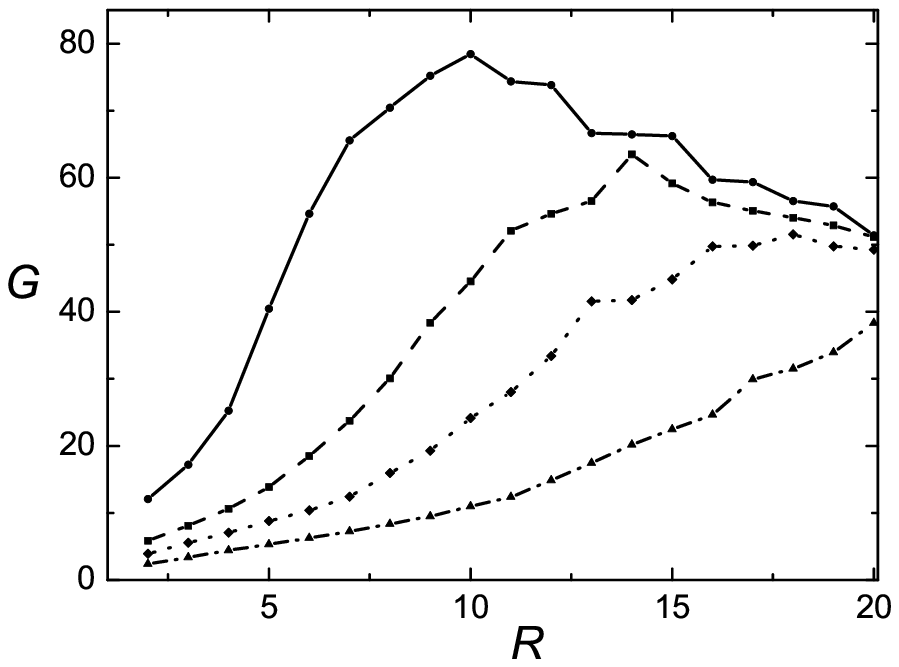}}}
\resizebox{9cm}{!}{\rotatebox[origin=c]{0}{\includegraphics{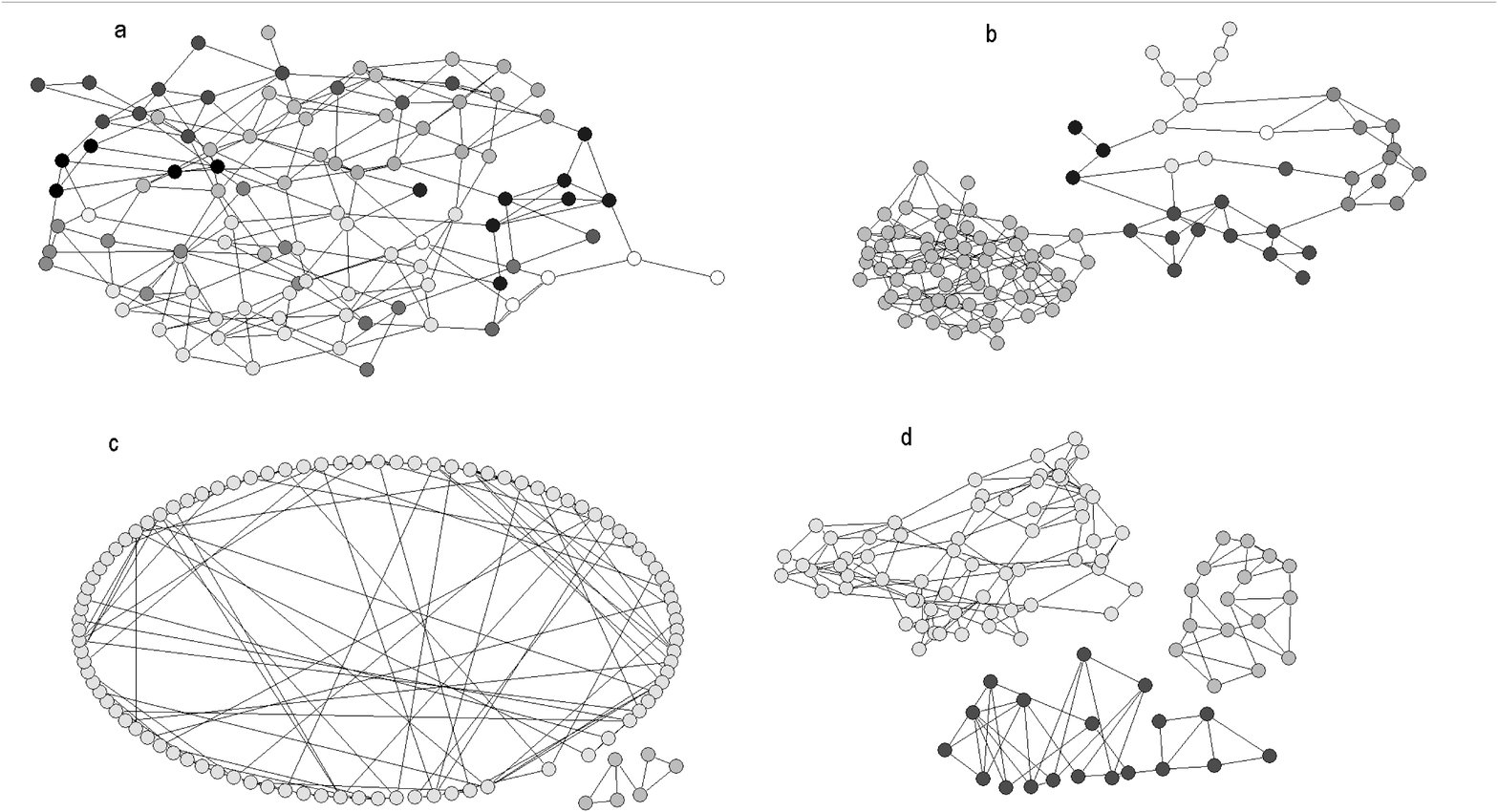}}}
\caption{$a$) Group size vs. R for different values of the vector
size: $q=10$(triangle), $q=20$ (diamond), $q=30$ (square), $q=50$
(circle). $b$) Final configurations of networks with variable
vector size. The results correspond to $q=50$, $t_r=0.1$, $t_n=10$
and  a)$R=2$, b) $R=5$; c) $R=15$, d)$R=20$}, \label{fig2}
\end{figure}

Plotted in Fig. \ref{fig2}-$a$ we observe $G$, the group size
versus $R$, with $q$ varying between 10 and 50. The results for
each value of $R$ is averaged over 500 realizations. As can be
seen in the figures displaying examples of final topologies, Fig.
\ref{fig2}-$b$, individuals sharing the same profile tend to
aggregate conforming communities within the population. The
figures correspond to only one realization  for $q=50$ and $R=2$,
5, 15 and 20 respectively. In these cases $t_r=0.1$, $t_n=10$

\begin{figure}[!h]
\centering
\resizebox{13cm}{!}{\rotatebox[origin=c]{0}{\includegraphics{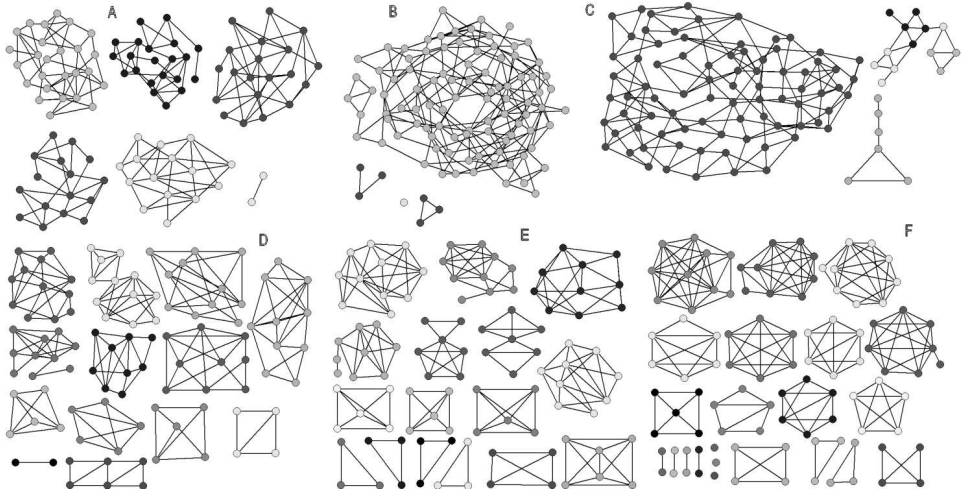}}}
\caption{ Final configurations of networks.  The results
correspond to $q=50$, $R=10$, and (a) $t_r=0.1$, $t_n=1$,(b)
$t_r=0.1$, $t_n=10$,(c) $t_r=0.1$, $t_n=100$,(d) $t_r=1$,
$t_n=1$,(e) $t_r=1$, $t_n=10$,(f) $t_r=1$, $t_n=100$ },
\label{newnet}
\end{figure}

When we  consider lower values of  $t_n$ we observe that the
behavior of the number of groups is similar to the one displayed
in Fig. \ref{staxbr}, but showing  a higher degree of segregation
for lower values of $R$. The segregation of the network is due to
the fact that in the last case we have let the convergence
behavior act less time on the system between two consecutive
changes in the topology produced by the network dynamics. If $t_r
\leq 1$ then the network is fragmented into small groups for any
of the values of $R$. On the contrary, if we increase the value of
$t_r$ we observe a rather robust behavior of the system with
similar final topologies. The reverse is true when changing the
values of $t_n$. This is due to the competition of two dynamics
with opposite effects. On one side, the repertoire dynamics based
on imitation drives the system to a homogeneous state. On the
other side, the network dynamics tends to segregate the network
and inhibit the convergence to uniformity. As an example, we show
in Fig. \ref{newnet} a typical outcome of individual realizations
with $R=q=20$, and different combinations of $t_r$ and $t_n$.

In the limiting case, for large values of $t_n$, we recover the
already know results for a static network \cite{palm2,kup,cast}.
In this case, the disorder of the underlying network promotes the
convergence to a homogeneous state whereas the increase of the
ratio $q/R$ facilitates the heterogeneity.

\subsection{Fixed  vector size}

\begin{figure}[!h]
\centering
\resizebox{9cm}{!}{\rotatebox[origin=c]{0}{\includegraphics{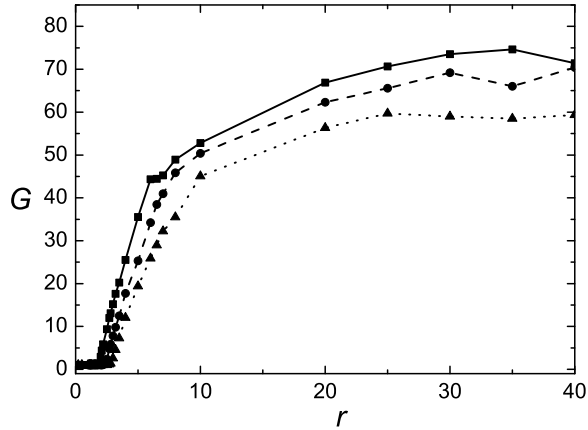}}}
\resizebox{9cm}{!}{\rotatebox[origin=c]{0}{\includegraphics{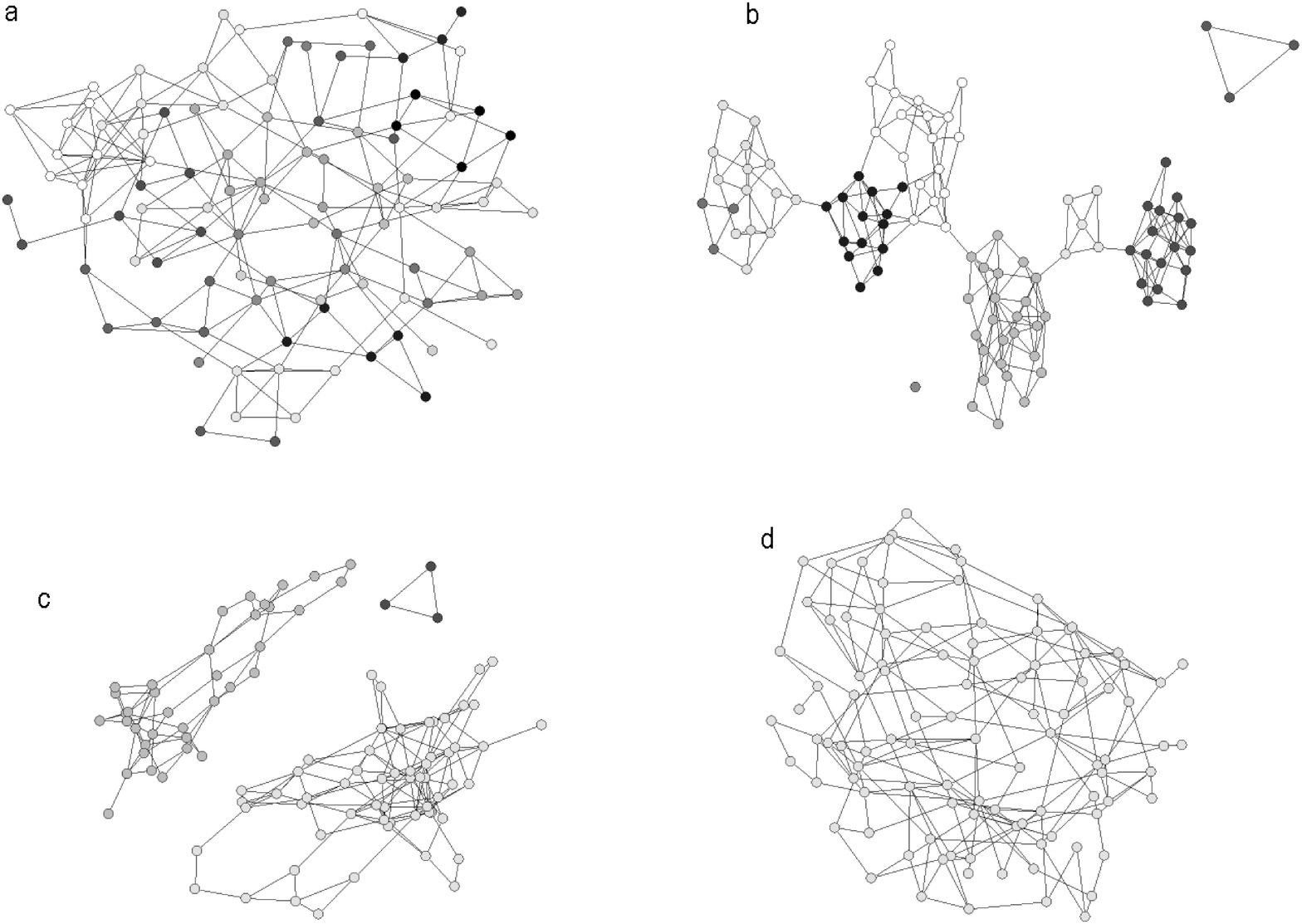}}}
\caption{$a$) Number of group $G$ vs. $r=\frac{1}{\chi}$ for
different values of $q$; $q=10$ (triangle), $q=20$ (circle) $q=40$
(square). $b$) Final network configurations for $q=20$ and
(a)$r=2$, (b)$r=5$, (c)$r=10$, (d)$r=25$. } \label{fig3}
\end{figure}

Throughout the realizations discussed in this section, we have
considered a fixed $R$ value, $R=20$. The net effect of a more
efficient communication is the convergence to more homogeneous
final configurations. The size of the groups increases as $\chi$
diminishes.  Figure \ref{fig3} shows the values of $G$ as a
function of $r=1/\chi$, for different values of $q$. It is clear
that the group size suddenly increases when crossing a given value
of $\chi$, i. e., when the interaction between individuals is more
probable, corresponding to $r\approx 2.5$. The most notable
difference with respect to the case previously discussed is that
the increment in group size is maintained. As it is clearly seen
in Fig. \ref{fig2}-$a$, the former case showed that the group size
non monotonic behaviour.

\subsection{Conclusions}

It has been nicely argued in \cite{oht} that cooperation is a
fundamental aspect of {\it all} biological system. In their paper,
they found that a {\it very simple and intuitive rule} is a good
approximation for all the stationary social structures analyzed.
In the present work, we argued that in social signaling systems
the emerging structure is the result of the coevolution of the
repertoire and the interaction between the elements of the system.
The repertory evolution can be thought as the lower level of
cooperation evolution since, in any system, for elements to
cooperate it should exist previously certain rules or information
that are understandable for all of them. Our results, based on
simple, but realistic, assumption of imitation and reconnection to
similar element of the system shows clearly that the
experimentally observed features in structured social behavior can
be understood following these basic rules. Also we explore another
biological aspect explicitly in our model, that is the relation
between the size of the social group and the cognitive abilities
of a given species. In doing that we make the connection between
the size of a given repertoire size, the structure of a given
social network (through the group size) and the cognitive
abilities of the elements that belong to it. As discussed in
\cite{Kudo2001}, the interplay between neocortex size and group
size is that coalitions allows animals to minimize the costs of
living in groups. By coalitions, we understand the organization of
a big population into smaller subgroups. As the cost of diary
activities, like travel cost or the increasing of day journey,
increase in groups of primates, they can manage, via coalition, to
decease such a negative affect of grouping. In this context, our
results shows that the conformation of smaller units in a
population is the results of a fast evolution of affinity in the
shared information, and that can be linked at the same time with
the capacities of compute it.

A fascinating aspect of this research is the connection with the
evolution of intelligence. Why intelligence evolve? or in other
words: How the ability to process information in a useful way
evolve? Our simple model is not a answer for that, but it has some
ingredients that are close to the social intelligence hypothesis
which state that intelligence evolve not to solve physical
problems but to process and use social information, such as
understanding the existing alliances and use this information for
deception \cite{rea,joll,wor}.

The detailed entanglement of the set of  signals that a social
system can process and the topology of the structure that the
described process generates are beyond the scope of this work. We
think that this hardly explored region of knowledge can bring some
light for understanding the fundamentals rules of social and
biological systems which can be uses as innovating tools for
practical issues in our social environments.

\section{Acknowledgment}
The author acknowledges support from  Agencia Nacional de
Promoci\'on Cient\'{\i}fica y Tecnol\'ogica (PICT 04/943), Consejo
Nacional de Investigaciones Cient\'{\i}ficas y T\'ecnicas (PIP
112-200801-00076), and Universidad Nacional de Cuyo (06/C304).

\end{document}